\documentclass[a4paper]{jpconf}
\usepackage{graphicx}
\usepackage{lineno}
\usepackage{atlasphysics} 
\usepackage{url}
\newcommand\mtTwo{\ensuremath{m_{\mathrm{T}2}}}

\begin{document}
\title{Experimental results on SUSY searches with top}

\author{Till Eifert,\\On behalf of the CMS and ATLAS Collaborations.}

\address{CERN, PH department, CH-1211 Geneva 23, Switzerland}

\ead{till.eifert@cern.ch}

\begin{abstract}
Searches for supersymmetric partner particles of the top and bottom quarks at the Large Hadron Collider are reviewed. The focus is on the status of searches for a relatively light partner of the top quark performed by the CMS and ATLAS Collaborations. No excess beyond Standard Model expectations is observed and exclusion limits are set on the masses of supersymmetric particles. 
\end{abstract}

\section{Introduction}
Supersymmetry (SUSY)~\cite{Golfand:1971iw} is widely considered to remain the best proposal for physics beyond the Standard Model (SM), resolving the main shortcomings of the SM while predicting an elementary Higgs scalar, with a SM-like limit and a mass below  $\sim135$\,\GeV.
Searches for `natural SUSY' models~\cite{Barbieri:1987fn} have received much attention in the past years. 
These models focus on those SUSY particles that are necessary to achieve a low-level of fine-tuning: light SUSY partners of the Higgs bosons (higgsinos) and top and bottom quarks (stop and sobttom, respectively), as well as a not-too-heavy partner of the gluon (gluino). Within the scope of strong production and $R$-parity conserving models, this corresponds to searches for pair production of 
stops, or sbottoms, or gluinos that decay via a stop or sbottom (gluino-mediated stop or sbottom production).\footnote{Searches for electroweak production of SUSY particles, and $R$-parity violating models are also performed at the LHC but not covered in this review.} Both CMS~\cite{CMS} and ATLAS~\cite{ATLAS} have performed extensive search programs for third generation squarks relying on the complementarity of many analyses as summarized in Table~\ref{overview}.

This review gives an overview of third generation analyses, with an emphasis on searches for a relatively light stop. 

\begin{center}
\begin{table}[h]
\centering
\begin{tabular}{l | l | l}
\hline
Search mode & \multicolumn{1}{c|}{CMS} & \multicolumn{1}{c}{ATLAS}\\
\hline
Direct stop & 0L~\cite{CMS_directstop_0L}, 1L~\cite{CMS_directstop_1L} & 0L~\cite{ATLAS_directstop_0L}, 1L~\cite{ATLAS_directstop_1L}, 2L~\cite{ATLAS_directstop_2L}\\
                 & combination of 0L razor and 1L~\cite{CMS_directstop_comb_razor_1L} & stop in charm~\cite{ATLAS_directstop_charm}\\
                 & stop in charm~\cite{CMS_directstop_charm}                   & heavier-stop in Z/GMBS stop~\cite{ATLAS_directstop2}\\
                 & heavier-stop in Z/h~\cite{CMS_directstop2}                   & stop in stau~\cite{ATLAS_directstop_stau}\\
\hline
Direct sbottom & 0L~\cite{CMS_directsbottom} & 0L~\cite{ATLAS_directsbottom}\\
\hline
Gluino-mediated stop & (0,1,2)L razor~\cite{CMS_gluinomed_razor} & (0,1)L, $\ge3$ $b$-jets~\cite{ATLAS_gluinomed_3b}\\
                           & 0L~\cite{CMS_gluinomed_0L}, 1L~\cite{CMS_gluinomed_1L} & 2 same-sign L / 3L~\cite{ATLAS_gluinomed_2SSL} \\
                           & 2 opposite-sign L~\cite{CMS_gluinomed_2OSL} & (also direct sbottom search)\\
                           & 2 same-sign L~\cite{CMS_gluinomed_2SSL},  3L~\cite{CMS_gluinomed_3L} &  0L, $\ge$(7--10) jets~\cite{ATLAS_gluinomed_multijets}  \\
\hline
Constraints on stop from: & & \ttbar\ cross section~\cite{ATLAS_top_cross_section}\\                           
  & & \ttbar\ spin correlation~\cite{ATLAS_top_spin_correlation}\\
\hline
\end{tabular}
\caption{\label{overview}Overview of CMS and ATLAS searches for third generation squarks. `L'  denotes an
isolated electron or muon, and `heavier-stop' refers to the heavier-stop mass eigenstate.
}
\end{table}
\end{center}

\section{Searches for third generation squarks}

\begin{figure}
\begin{center}
\includegraphics[width=0.49\textwidth]{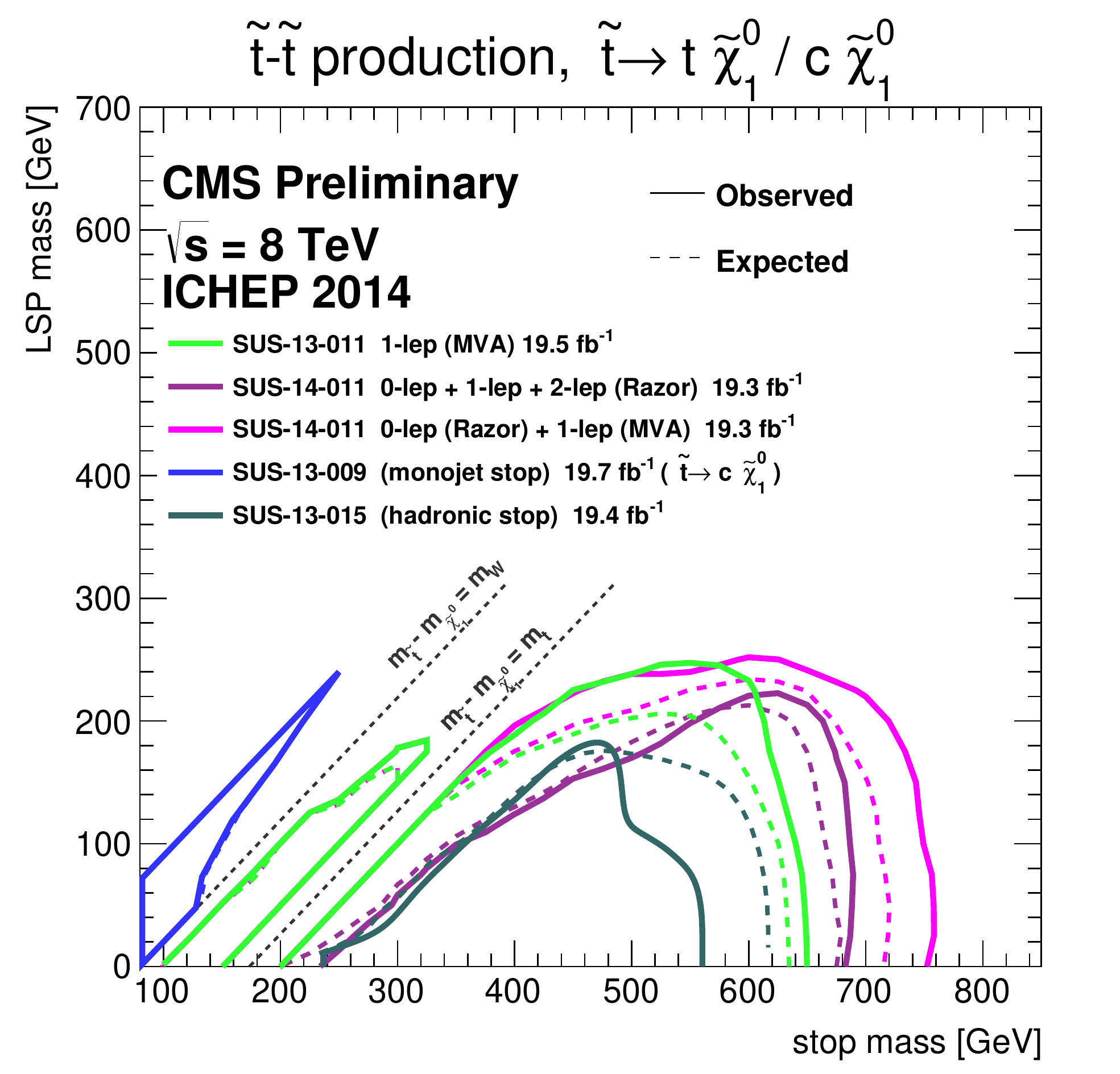} 
\includegraphics[width=0.49\textwidth]{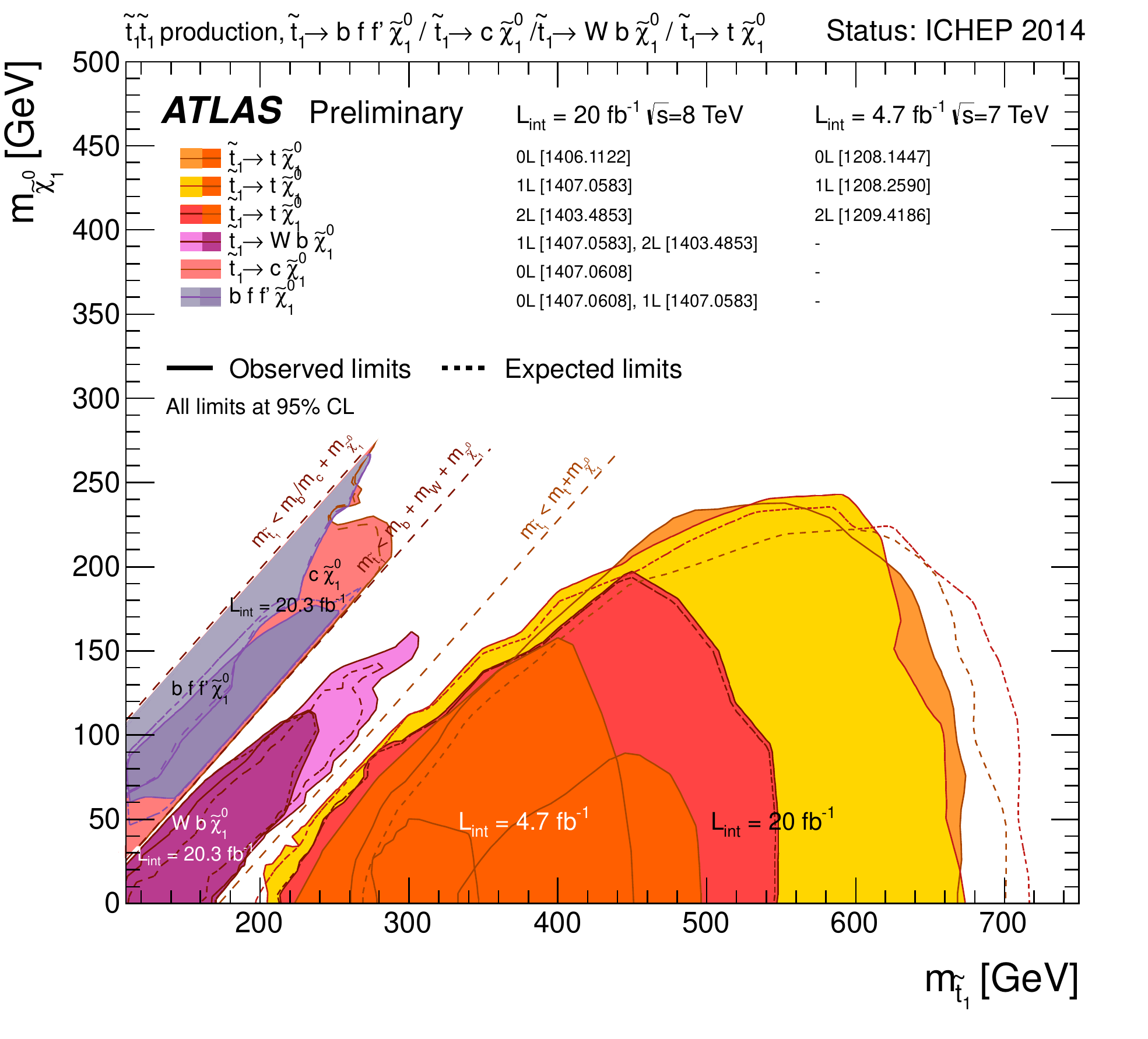} 
\caption{LHC stop exclusion limits at 95\% CL for the shown decay modes, assuming a branching fraction of 100\% in each case~\cite{CMS_directstop_0L,CMS_directstop_1L,ATLAS_directstop_0L,ATLAS_directstop_1L,ATLAS_directstop_2L,CMS_directstop_comb_razor_1L,ATLAS_directstop_charm,CMS_directstop_charm}. 
\label{fig:stop_overview}}
\end{center}
\end{figure}

Figure~\ref{fig:stop_overview} shows a summary of the exclusion limits obtained from the LHC direct stop searches listed in table~\ref{overview}.
The stop decay mode depends, amongst other things, on the stop mass itself and the mass spectrum of other SUSY particles. Assuming the stop is the next-to-lightest SUSY particle, its open decay modes depend on the available phase space:  to a top quark and the lightest SUSY particle (LSP); or to a bottom quark, a $W$ boson, and the LSP (three-body decay via off-shell top quark); or to a bottom quark, an off-shell $W$ boson, and the LSP (four-body decay); or via loop suppressed diagrams, for example to a charm quark and the LSP.  All of these decay modes have been considered by the LHC searches. 
The reach up to around 700\,\GeV\ in stop mass is limited by the cross section, while the gaps at lower stop masses arise due to experimental challenges in separating the stop signal from the main background, namely top quark pair production (\ttbar). 


A wide range of experimental techniques is employed in the search programs. The very compressed scenario, where the stop mass is only slightly above the mass of its decay products (of the four-body process, or the decay to a charm quark and the LSP) approached using a mono-jet search that requires a jet from initial-state radiation. The search for the four-body decay with a slightly less compressed mass spectrum takes advantage of selecting low-momentum leptons (soft-lepton analysis), while the search for the decay via a charm quark relies on charm-tagging.  The three-body decay searches are performed in the 1L and 2L channels\footnote{The letter `L'  denotes an isolated electron or muon.} with dedicated selections, exploiting features in kinematic variables (such as \mtTwo~\cite{mT21}) or using multi-variate analysis (MVA) techniques. The diagonal gap between the three-body and the two-body decays is discussed in the next section.  The search sensitivity for a stop with a mass of 600--700\,\GeV\ is dominated by the 0L and 1L channels. Both channels (in ATLAS) use selections with large-radius jets above some mass and momentum thresholds to more efficiently collect the collimated decay products of boosted objects (top quark, $W$ boson).

The dominant background tends to arise from \ttbar\ production. For the 0L (1L) channels, the selection of large missing transverse momentum and, for the 1L channel, of a large transverse mass well above the $W$ boson mass reduce the fully-hadronic (semi-leptonic) \ttbar\ components considerably. As a result, the semi-leptonic (di-leptonic) \ttbar\ component becomes the dominant background in the 0L (1L) search channels.  A number of techniques and tools have been developed and employed to reduce this dominant background, including: dedicated \mtTwo-like variables,  reconstruction of the hadronically decaying top quark, hadronic-tau vetos.

\section{Stealth stop}

One particularly interesting exclusion gap in the stop--LSP mass plane is the so-called `stealth' stop scenario, where the stop mass is close to the mass of the top quark and the LSP. 
The resulting kinematic properties of such a stop model closely resemble those of \ttbar\ events, while the stop pair production cross section is only about 15\% with respect to that of \ttbar\ production (for a stop mass similar to the top quark mass). 
Separating signal from background relies on small differences in the kinematic properties, induced by the effects of spin differences and the residual momentum in the two LSPs. 
The use of multi-bin (ATLAS) and MVA (CMS) techniques has pushed the search exclusion sensitivities to approach the top quark mass from above as close as  $25$\,\GeV. 

\begin{figure}
\begin{center}
\includegraphics[width=0.49\textwidth]{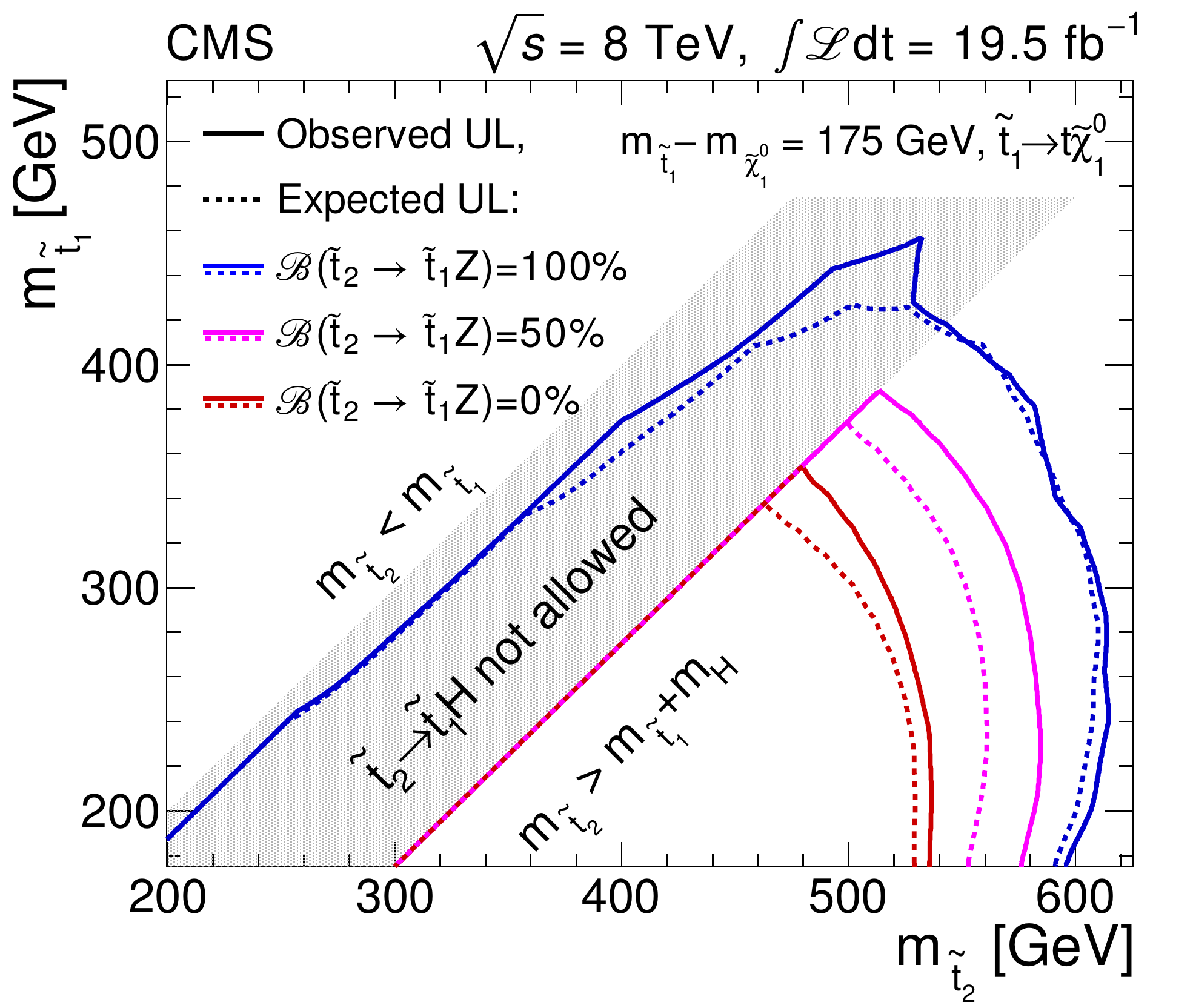} 
\includegraphics[width=0.49\textwidth]{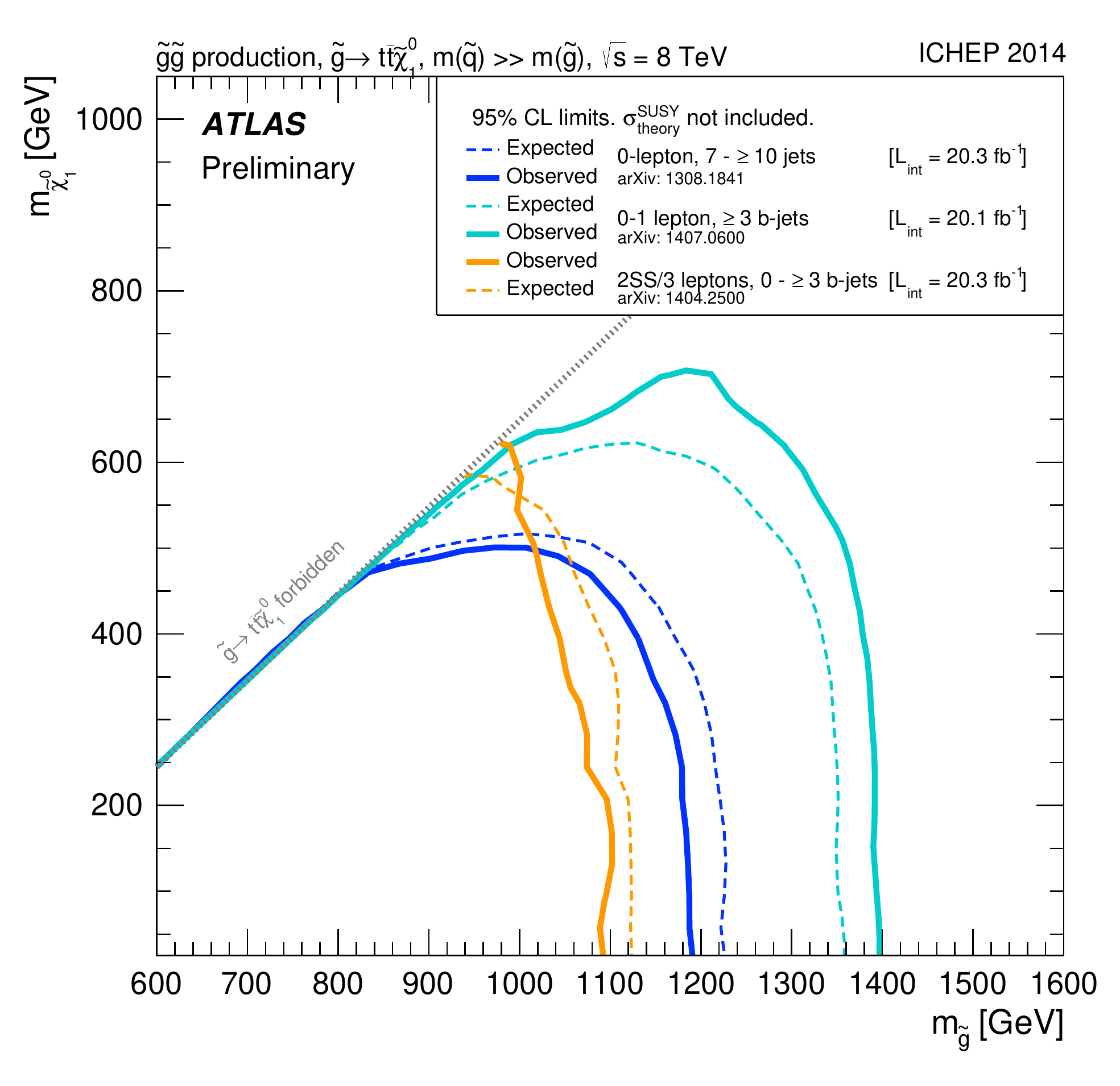} 
\caption{Exclusion limits from searches for the heavier-stop mass eigenstate~\cite{CMS_directstop2} (left), and for gluino-mediated stop production~\cite{ATLAS_gluinomed_3b} (right). Similar results are also available from the other experiment in both cases.
\label{fig:stealth_stop}}
\end{center}
\end{figure}

A different approach is to exploit searches for heavier SUSY particles, where the heavier SUSY particle is assumed to decay via a (stealth) stop. 
Figure~\ref{fig:stealth_stop} (left plot) shows the exclusion limits from CMS searches~\cite{CMS_directstop2} for the heavier-stop mass eigenstate (\stoptwo). The \stoptwo\ is assumed to decay to the lighter-stop mass eigenstate either
via a Higgs boson or a $Z$ boson. The searches exploit various experimental channels with 1L, 2L (same-sign or opposite-sign), and $\ge 3$L, as well as up to 4 $b$-tags, and selections of on- or off-shell $Z$ boson.
The 3L channel dominates the search sensitivity, and the limits reach up to about 600\,\GeV\ in \stoptwo\ mass, assuming a stealth lighter-stop mass eigenstate. ATLAS has published similar exclusion limits~\cite{ATLAS_directstop2}, but not yet considering the decay mode via a Higgs boson.

ATLAS search limits for gluino pair production where each gluino is assumed to decay via a stop to two top quarks and the LSP are shown in figure~\ref{fig:stealth_stop} (right plot). CMS has similar results.  The decay to four top quarks and extra missing momentum is a spectacular signature. The limits at high gluino mass are dominated by the (0,1)L, $\ge3$ $b$-jets search~\cite{ATLAS_gluinomed_3b}, where one of the experimental challenges is the estimation of both the irreducible and reducible backgrounds with $\ge 3$ $b$-jets. 
A gluino is excluded up to a mass of about 1.4\,\TeV. This limit is only weakly dependent on the stop mass.

\section{Stop constraints from \ttbar\ measurements}

\begin{figure}
\begin{center}
\includegraphics[width=0.49\textwidth]{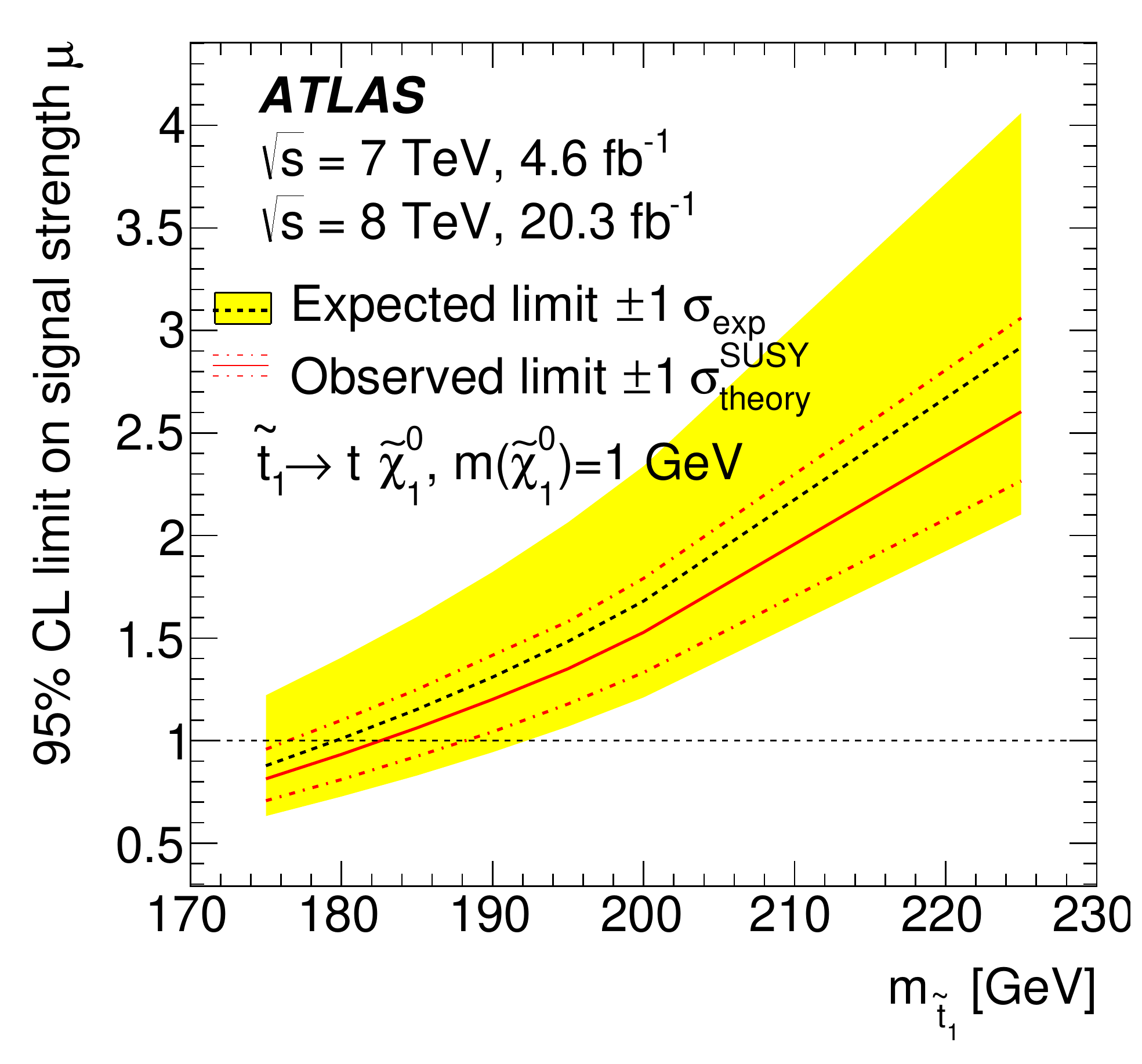} 
\includegraphics[width=0.49\textwidth]{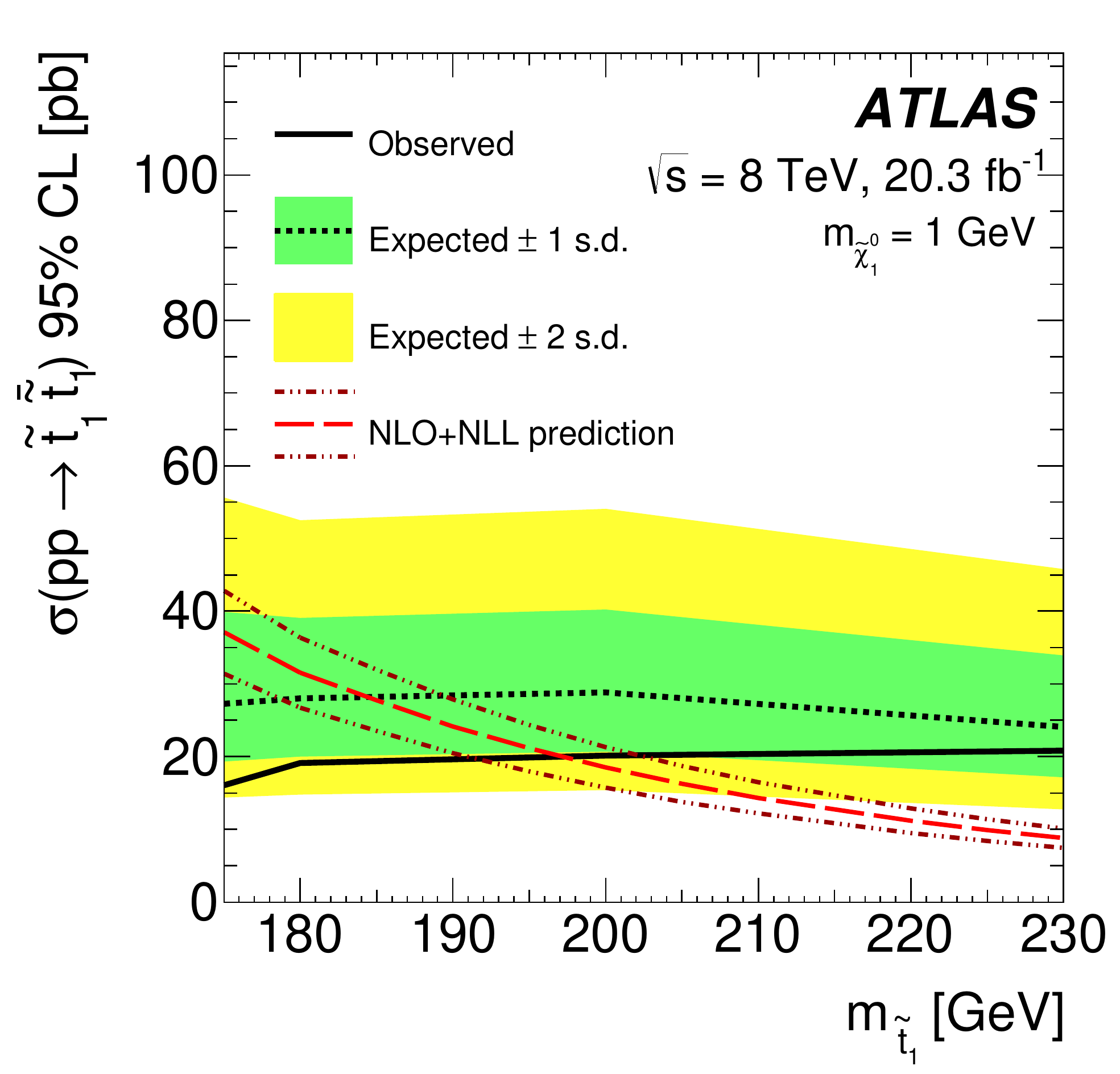} 
\caption{Stop exclusion limits from ATLAS \ttbar\ measurements of the cross section~\cite{ATLAS_top_cross_section} (left)
 and spin correlation~\cite{ATLAS_top_spin_correlation} (right).
\label{fig:top}}
\end{center}
\end{figure}

SM top quark measurements can be exploited to get a handle on the potential presence of a stealth stop. 
The ATLAS \ttbar\ cross section measurement~\cite{ATLAS_top_cross_section} 
selects electron-muon events, and takes the ratio of events with exactly one $b$-tagged and exactly two $b$-tagged jets
to reduce systematic uncertainties. The resulting total uncertainty is about 4\%,
which together with the state-of-the-art theoretical predictions (with an uncertainty of about 5\%) provides
some sensitivity to the $\mathcal{O}(10\%)$ contribution of a stealth stop.\footnote{See ref.~\cite{sneakyStop} for a discussion of the potential bias on the top-quark mass.} Figure~\ref{fig:top} (left plot) shows the resulting 
exclusion limit as a function of the stop mass, setting a 95\% CL exclusion limit between the top threshold and 177\,\GeV. This is assuming that the stop decays with 100\% branching fraction to a top quark and the LSP, and that the LSP mass is 1\,\GeV, 
and that the stop is mostly partner of the right-handed top quark. 

The precise measurement of the \ttbar\ spin correlation is sensitive to the presence of a stealth stop due to the spin-0 nature of the stop~\cite{spin-correlation}. 
The ATLAS measurement~\cite{ATLAS_top_spin_correlation} selects di-lepton events and requires at least two jets with at least one $b$-tag as well as missing transverse momentum to suppress backgrounds. 
The spin correlation measurement is based on the shape of the azimuthal angle between the two leptons ($\Delta \phi$). 
Figure~\ref{fig:top} (right plot) shows the resulting 95\% CL exclusion limit as a function of the stop mass, extending the limit up to 191\,\GeV. 
This is based on the same stop model as above.  Without the shape information in the $\Delta \phi$ distribution, the limit would deteriorate by 30--40\%. 
CMS has not yet used \ttbar\ measurements to search for, and set limits on a stealth stop. 

\section{Summary and outlook}
Both CMS and ATLAS have performed extensive search programs for the stop (SUSY partner of the top quark), covering a wide range of decay modes and stop mass space. 
Exclusion limits reach up to a stop mass of about 700\,\GeV, but leaving some gaps behind. A particularly interesting gap is the stealth stop scenario, where the stop mass is close to the top quark mass. Top quark measurements in ATLAS have been used to set first exclusion limits in this stealth stop scenario.

CMS and ATLAS have studied the prospects for the (HL) LHC with 300 and 3000\,\ifb~\cite{CMS-prospects, ATLAS-prospects}. Searches for gluino-mediated stop production are expected to reach gluino masses beyond 2\,\TeV, while searches for direct stop production are expected to have a reach well beyond 1\,\TeV.

\end{document}